\documentclass[10pt,letterpaper,amsmath,amsfonts,amssymb,aps,pra,twocolumn,superscriptaddress]{revtex4-1}
\usepackage[latin1]{inputenc}
\usepackage{bm}
\usepackage[pdftex]{graphicx}
\usepackage{color}
\usepackage{verbatim}
\usepackage{xargs}[2008/03/08]
\usepackage[colorinlistoftodos]{todonotes}
\usepackage{amsmath}
\usepackage{amsfonts}
\usepackage{amssymb}
\usepackage{bm}
\usepackage[pdftex]{graphicx}
\usepackage{etoolbox}
\newcommand{\op}[1]{\hat{\bm #1}}
\newcommand{\ket}[1]{\lvert #1\rangle}
\newcommand{\bra}[1]{\langle #1 \rvert}
\newcommand{\pr}[1]{\ket{#1}\bra{#1}}
\newcommand{\ipr}[2]{\langle #1 \vert #2 \rangle}
\newcommand{\mean}[1]{\left\langle #1 \right\rangle}

\begin{document}
\title{Noise suppression in inverse weak value based phase detection}
\author{Kevin Lyons}
\affiliation{Department of Physics and Astronomy, University of Rochester, Rochester, New York 14627, USA}
\affiliation{Center for Coherence and Quantum Optics, University of Rochester, Rochester, New York 14627, USA}
\author{John C. Howell}
\author{Andrew N. Jordan}
\affiliation{Department of Physics and Astronomy, University of Rochester, Rochester, New York 14627, USA}
\affiliation{Center for Coherence and Quantum Optics, University of Rochester, Rochester, New York 14627, USA}
\affiliation{Institute for Quantum Studies, Chapman University, 1 University Drive, Orange, CA 92866, USA}

\date{\today}

\begin{abstract}
We examine the effect of different sources of technical noise on inverse weak value-based precision phase measurements.  We find that this type of measurement is similarly robust to technical noise as related experiments in the weak value regime.  In particular, the measurements considered here are robust to additive Gaussian white noise and angular jitter noise commonly encountered in optical experiments.  Additionally, we show the same techniques used for precision phase measurement can be used with the same technical advantages for optical frequency measurements.   
\end{abstract}

\maketitle

\section{Introduction}
In the original work on weak value measurement by Aharonov, Albert, and Vaidman \cite{Aharonov1988}, the authors showed that the shift of a ``meter'' state (\emph{i.e.}~a pointer state), which is entangled with a ``system'' state, can be amplified arbitrarily in some basis at the cost of reducing the probability of detecting a given event due to a process termed postselection.  This amplification led eventually to interest in using weak values for the purpose of precision measurements \cite{Hosten2008,Dixon2009}. 
 
In recent years, postselected weak measurements have been used to great effect in metrological applications \cite{Dixon2009,Starling2009,Starling2010,Starling2010a,Howell2010,Hogan:11,Pfeifer:11,Egan:12,PhysRevLett.109.013901,Zhou2012,Strubi2013,Viza:13,PhysRevLettv111p033604,Zhou2013,PhysRevLett.112.200401,PhysRevA.89.012126,salazar2015enhancement,Martinez-Rincon:17}.  For a more complete overview of both the relevant theory and experiments, see the review article by Dressel \emph{et al.}~\cite{Dressel2014}.  There has been recent theoretical work 
examining the use of quantum optical resources in the meter degree of freedom which shows that it is possible for a postselected measurement to achieve higher sensitivity than non-postselected measurements \cite{Pang2014,Pang2015}.  However, the overwhelming majority of experimental work to date has used classical (\emph{i.e.}~coherent) states which can not improve over traditional schemes with postselection in ideal, shot-noise limited systems \cite{Pang2015}.

The principal reason weak measurements have improved sensitivity is that they allow certain types of technical  noise or other experimental limitations to be overcome, while still achieving the same sensitivity as ideal traditional measurement schemes \cite{Starling2009,Jordan2014,Knee2014,viza2015experimentally,PhysRevA.91.062107,torres2016weak,harris2016weak}.  Previous theoretical work demonstrating the effectiveness of weak value amplification in reducing the negative effects of technical noise has been in what we term the \emph{weak value regime}.  Namely, a small but known phase parameter is used in the system to amplify a very small and unknown interaction parameter (\emph{i.e.,}~one coupling system and meter quantum dynamical variables).  It has also been recently proven that in the case of systematic noise on the meter, weak value amplification can suppress this systematic contribution in comparison to a direct measurement method \cite{Pang2016}.  Further improvements in precision can be obtained by recycling the non-selected events \cite{dressel2013strengthening,lyons2015power,byard2015pulse,wang2016experimental}.

In this work we consider the opposite case where we have a small but known interaction parameter which is used to amplify a very small and unknown phase.  We refer to this as the \emph{inverse weak value regime}, following previous work \cite{Starling2010a,Dressel2013,Lyons2015,Kofman2012}. 

The paper is organized as follows: in Sec.~\ref{sec:inverse weak values}, we briefly review weak measurements in a two level system, and compare basic results in the weak value and inverse weak value regimes.  In Sec.~\ref{sec:noiseless}, we calculate the Fisher information for weak measurements in the inverse weak value regime, and compare the result with traditional phase estimation schemes. In Sec.~\ref{sec:gaussian noise}, we treat the case of uncorrelated additive Gaussian technical noise, and in Sec.~\ref{sec:jitter} we treat angular jitter and diffraction effects in optical phase measurements.  In Sec.~\ref{sec:frequency}, we extend our analysis to the case of precision frequency measurements in the inverse weak value regime, and demonstrate that such a measurement will have the same robustness to technical noise as phase measurements.

\section{Inverse weak value}\label{sec:inverse weak values}
Here we briefly recall the usual weak value amplification process in a two level system, and then discuss how this relates to the process of amplification by the so-called inverse weak value.  

We take an initial ``system'' state $\ket{i}$ and ``meter'' state $\ket{\varphi}$ to be in a product state, \emph{i.e.,} $\ket{\Psi} = \ket{i} \otimes \ket{\varphi}$.  An effective interaction Hamiltonian 
\begin{align}\label{eq:kick hamiltonian}
	H_1 &= g_1(t) \op{\sigma}_z \otimes \op{x},
\end{align}
entangles the system and meter weakly, with $g(t) \ll 1$ for some short time interval.  Here $\op{\sigma}_z$ is the usual Pauli z spin operator which acts on the system state space, and $\op{x}$ is taken to be the (transverse) position operator acting on the meter state space.  If the intended system postselection state $\ket{f}$ is orthogonal to $\ket{i}$, it is also necessary to introduce a second interaction in order to create some small but nontrivial transition probability between the initial and final system states.  A simple choice would be 
\begin{align}\label{eq:phase hamiltonian}
	H_2 &= g_2(t) \op{\sigma}_z \otimes \op{1},
\end{align}
which only has a nontrivial action on the system space.  Qualitatively, we see that $g_1(t)$ generates a displacement in the basis of the operator conjugate to $\op{x}$, with the sign of the displacement opposite for each of the two basis vectors, and $g_2(t)$ generates a relative phase between the two basis vectors of $\op{\sigma}_z$.  The action of these two operators on the state $\ket{\Psi}$ is straightforwardly given by the unitary operators 
\begin{align}
	\nonumber\op{U}_1 &= \exp(-i k \op{\sigma}_z \otimes \op{x}), \\
	\op{U}_2 &= \exp(-i  \frac{\phi}{2} \op{\sigma}_z \otimes \op{1}),
\end{align}
where $k$ and $\phi$ are effective interaction parameters found by integrating the above Hamiltonians over the relevant time interval.  

Since $[\op{U}_1, \op{U}_2] = 0$, we can simply combine the operators as $\op{U} \equiv \op{U}_1 \op{U}_2$ by adding the exponents.

Beginning with our state $\ket{\Psi}$, we find that in the diagonal basis of $\op{x}$, we can represent the postselected state as 
\begin{align}\label{eq: general postselected state}
	\nonumber \ipr{x}{\Psi^{'}} &= \bra{x} \otimes \frac{\pr{f}}{\sqrt{p_f}} \op{U} \ket{i} \otimes \ket{\varphi}, \\
	&= \frac{-i \bra{f}\op{\sigma}_z \ket{i}}{\sqrt{p_f}} \sin \left(k x + \frac{\phi}{2}\right) \ipr{x}{\varphi} \ket{f},
\end{align}
where $p_f$ is the postselection probability.  From this point we may ignore the system state and focus on the meter after postselction.  The postelection probability is calculated directly as 
\begin{align}
	p_f &= \int_{-\infty}^{\infty} \text{d}x~ |\ipr{x}{\varphi}|^2 \left|\bra{f}\op{U} \ket{i}\right|^2.
\end{align} 
In the specific case of a Gaussian initial meter state with variance $\sigma^2$, this yields 
\begin{align}\label{eq:postselection probability}
	p_f &= \frac{1 -  e^{-2k^2\sigma^2}\cos\phi}{2}.
\end{align}
In the \emph{weak value regime}, we have $k\sigma$ small compared to $\frac{\phi}{2}$ for values of $x$ where $|\ipr{x}{\varphi}|$ is not very small.  For a concrete example we take $\ket{i} = 2^{-1/2}(\ket{0} + \ket{1})$, and  $\ket{f} = 2^{-1/2}(\ket{0} - \ket{1})$.  In the weak value regime this yields $p_f \approx |\bra{f}\op{U_2}\ket{i}|^2 =\sin^2 \frac{\phi}{2}$, and $\mean{x} \approx k \sigma^2 \cot \frac{\phi}{2}$.  Hence, a small known phase parameter $\phi/2$ is used to amplify $\mean{x}$, which in turn allows for an improved estimation of the unknown parameter $k$.

In the \emph{inverse weak value regime}, the situation is reversed, where one wishes to use a small known parameter $k$ to amplify an unknown phase parameter $\phi/{2}$.  This does not change Eq.~\eqref{eq: general postselected state}, but there is no simple way to calculate the postselection probability without assuming a particular initial meter state.  Also note that in this regime, there is certain to be an eigenvalue $x$ which yields zero amplitude upon postselection, \emph{i.e.,} where $k x + \phi/{2} = 0$.  Taking the same initial and final system states as above, and again assuming a Gaussian initial meter state with variance $\sigma^2$, we find 
\begin{align}\label{eq:noiseless iwv}
	\nonumber p_f &\approx k^2 \sigma^2, \\
	\mean{x} &\approx \frac{\phi}{k}.
\end{align}

\section{Fisher information}\label{sec:noiseless}
We now turn our attention to estimating the unknown parameter $\phi$ in the inverse weak value regime for a Gaussian initial meter state.  Hence, postselected events will be sampled from the probability distribution
\begin{align}\label{eq: general distribution}
	p(x|\phi) &= \frac{\left|\bra{f}\op{\sigma}_z \ket{i}\right|^2}{p_f}\sin^2\left(kx + \phi/2\right)\left|\ipr{x}{\varphi}\right|^2,
\end{align}  
where the postselection probability $p_f$ is given by Eq.~\eqref{eq:postselection probability}.
As in the previous section, we will choose orthogonal initial and final states so that $\bra{f}\op{\sigma}_z\ket{i} = 1$.
The metric we will use to determine the sensitivity of the estimation scheme is the Fisher information, defined by 
\begin{align}\label{eq:noiseless fisher}
	\nonumber \mathcal{I}(\phi) &= -\int_{-\infty}^{\infty} \text{d}x~ p(x|\phi)\partial_\phi^2 \ln p(x|\phi), \\
	&= \frac{e^{4k^2\sigma^2} - 1}{(e^{2k^2\sigma^2}-\cos \phi)^2}.
\end{align}

If there are $\nu$ total events, then $p_f \nu$ events will be postselected.  Since the Fisher information scales linearly with the number of independent detected events, we are interested in the quantity
\begin{align}
	\mathcal{I}_\nu (\phi) &= p_f \nu \mathcal{I}(\phi), \nonumber \\
	&\approx  \frac{\nu}{2}\left(1 + e^{-2k^2\sigma^2}\right),
\end{align} 
where the approximation is from expanding to first order in $\phi$ (\emph{i.e.}~$\cos \phi \approx 1$). This expression approaches $\nu$ for small $k\sigma$.  Since $\mathcal{I}_\nu(\phi) = \nu$ for more conventional phase estimation schemes such as balanced homodyne detection (or any scheme with Poissonian uncertainties), the inverse weak value approach can approximately recover the full Fisher information with a small subensemble of all events.  The same result has been found in the weak value regime \cite{Jordan2014}.

\section{Uncorrelated gaussian noise}\label{sec:gaussian noise}
\begin{figure}
	\includegraphics[]{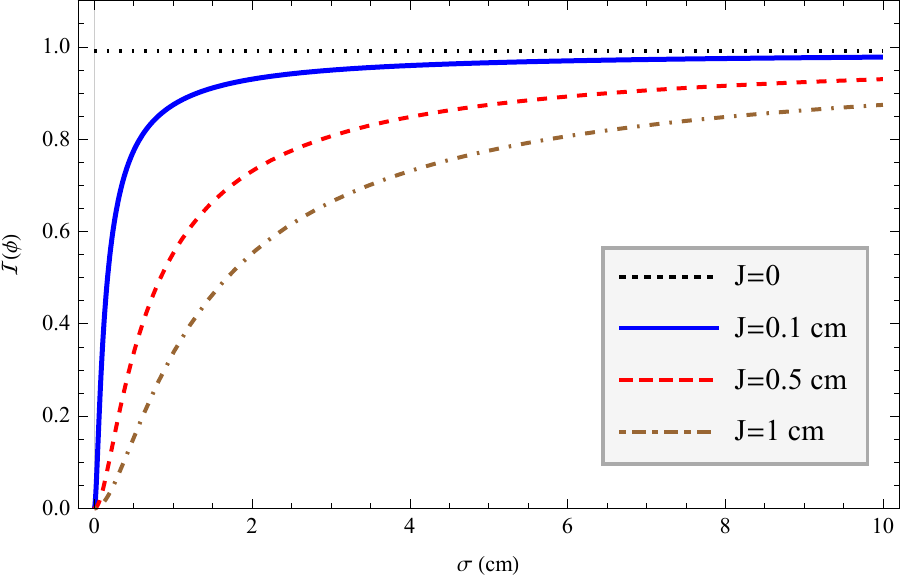}
	\caption{Fisher information for estimation of the parameter $\phi$ under the effect of additive Gaussian white noise, for different noise amplitudes $J$ as a function of the initial state standard deviation $\sigma$.  The product $k\sigma$ is held constant in this plot to keep the postselection probability \eqref{eq:postselection probability} constant.  }
\end{figure}
For the case of uncorrelated Gaussian technical noise, such as might be the case with beam jitter in an optical experiment, the measured signal position $s$ will be the sum of a position $x$ drawn from the distribution \eqref{eq: general distribution} and a random displacement $\xi$ drawn from a Gaussian distribution with zero mean and variance $J^2$.  Hence, $s = x + \xi$, and 
\begin{align}\label{eq:gaussian noise distribution}
	p(s|\phi) &= \frac{1}{\sqrt{2\pi J^2}}\int \text{d}\xi\int \text{d}x~e^{-\xi^2/2J^2}p(x|\phi)\delta(s - x - \xi), \nonumber\\
	&= \frac{1}{\sqrt{2\pi J^2}}\int \text{d}x~\exp\left(\frac{-(s - x)^2}{2J^2}\right)p(x|\phi), \nonumber \\
	&\nonumber= \frac{1}{\sqrt{2\pi (J^2 + \sigma^2)}}\exp\left[\frac{-s^2}{2(J^2+\sigma^2)}\right]\left(1 - \right.\\ &\left.\cos\left[\frac{2ks\sigma^2}{J^2 + \sigma^2} + \phi \right]\exp\left[\frac{-2J^2k^2\sigma^2}{J^2 + \sigma^2}\right]\right).
\end{align}

It is difficult to directly calculate the Fisher information of this distribution for an estimator of the parameter $\phi$.  If we expand to linear order in $\phi$ and $k\sigma$ with $\phi < k\sigma < 1$ however, the integration is straightforward.  For clarity of the final result we also expand to first order in $J/\sigma$, however our conclusions are not dependent on this truncation.  The resulting information is given by
\begin{align}
	\mathcal{I}(\phi) &= \frac{1}{k^2\sigma^2}\left(1 - \sqrt{\frac{\pi}{2}}\frac{J}{\sigma}\right).
\end{align}
If we choose $\sigma \gg J$, this expression tends to 
\begin{align}
	\mathcal{I}(\phi) &= \frac{1}{k^2\sigma^2}.
\end{align}
Recall $k^2\sigma^2$ is equal to the postselection probability \eqref{eq:noiseless iwv} in the inverse weak value regime.  We also note this is equal to Eq.\eqref{eq:noiseless fisher} when expanded to lowest nonvanishing order in $\phi$ and $k\sigma$.  Hence, for a sufficiently large choice of $\sigma$ the Fisher information for $\nu$ independent events is given by 
\begin{align}
	\mathcal{I}_\nu(\phi) &= p_f \nu \mathcal{I}(\phi), \nonumber \\
	&\approx \nu.
\end{align}
Remarkably, it is possible to approximately recover the full Fisher information in the presence of additive Gaussian technical noise, again using only a small subensemble of total events, provided the Gaussian width $\sigma$ is sufficiently large.

\section{Angular jitter and diffraction}\label{sec:jitter}
\begin{figure}
	\includegraphics[scale=.6]{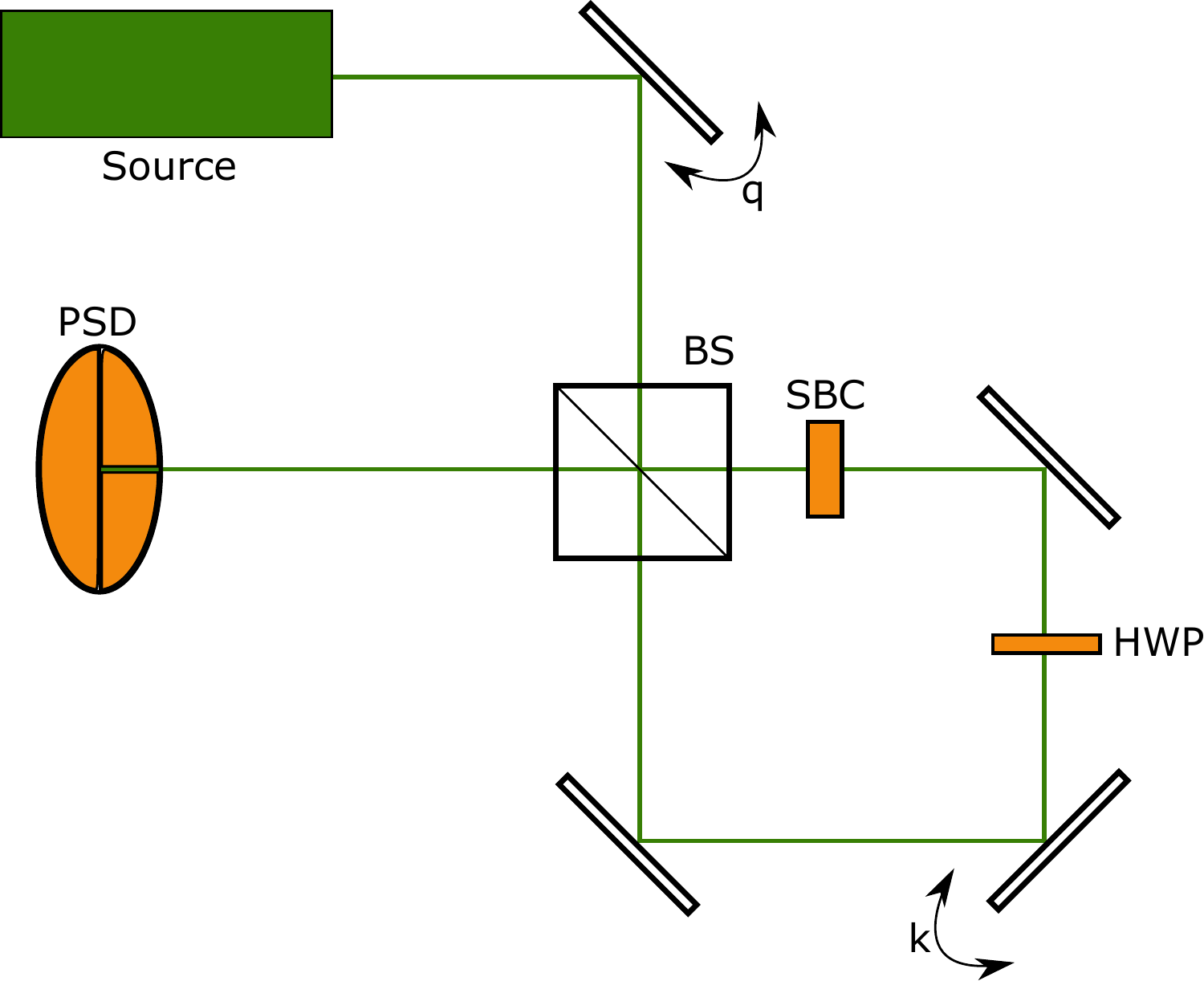}
	\caption{An optical two-level weak value amplification setup.  Angular jitter noise is modeled by the first mirror which imparts a random momentum shift $q$ to the beam.  The beam then passes through a beam splitter (BS) and acquires a path-dependent phase at the Soleil-Babinet compensator (SBC) with half-wave plate (HWP), and a path-dependent momentum shift $k$ at the titled signal mirror inside the Sagnac interferometer.  Postselected light is measured at the position sensing detector to determine the unknown phase shift $\phi$ imparted by the SBC.  The distance between the source and the symmetric point of the interferometer is $\ell_1$ and the distance between the symmetric point and the position-sensitive detector (PSD) is $\ell_2$.}\label{fig:jitter setup}
\end{figure}
\begin{figure}
	\includegraphics[scale=1]{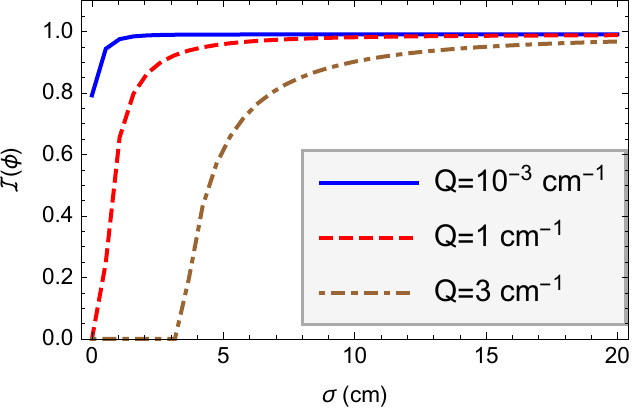}
	\caption{Fisher information as a function of beam width for two different amplitudes of angular jitter, with $k\sigma$ held at a constant value of $.1$.  The numeric values of the parameters are $\phi=10^{-3}$, $k_0 = 10^5 cm^{-1}$, $\ell_1 = 10^5 cm$, $\ell_2 = 10^2 cm$.  Evidently larger choices of $\sigma$ allow one to overcome angular jitter, even when $k \ll Q$.}\label{fig:jitter plot}
\end{figure}

We now consider propagation effects (\emph{e.g.} diffraction), and an angular jitter of an optical beam which we model as an additional tilted mirror as in Fig.~\ref{fig:jitter setup}.  We can model the unnormalized state at the detector as 
\begin{align}\label{eq:jitter state}
	\ipr{x}{\varphi} &= \bra{x}\op{U}_{\ell_2} \op{U}_{k} \op{U}_\phi \op{U}_{\ell_1} \op{U}_q \ket{\varphi_0},\nonumber \\
	&= \bra{x} e^{-i\op{p}^2\ell_2/2k_0}\sin(k\op{x} + \phi/2)e^{-i\op{p}^2\ell_1/2k_0}e^{iq\op{x}}\ket{\varphi_0},
\end{align}
where $\op{U}_\ell$ is a quadratic phase in momentum space which describes propagation of the beam front by a distance $\ell$, and $\ket{\varphi_0}$ is the initial Gaussian meter state.  If we take the angular jitter to be normally distributed with zero mean and variance $Q^2$, the overall distribution is given by 
\begin{align}
	p(x,q|\phi) &= \mathcal{N}\exp\left(\frac{-q^2}{2Q^2}\right)\left|\ipr{x}{\varphi}\right|^2,
\end{align}   
where $\mathcal{N}$ is a normalization constant.  Note that in the limit of a collimated beam (\emph{i.e.,} $\ell_1,\ell_2 \rightarrow 0$), the momentum kick $q$ will have no effect on the measured profile as it only appears as an overall phase factor in the wavefunction.  The Fisher information for an estimator of the parameter $\phi$ is then
\begin{align}
	\mathcal{I}(\phi) = -\int \text{d}x~ \text{d}q~ p(x,q|\phi)\partial_\phi^2 \ln p(x,q|\phi).
\end{align}
This integral is very complicated for the general state \eqref{eq:jitter state}, but becomes tractable with a few simplifying assumptions.  Namely, we expand to linear order in $\phi$ and $k\sigma$ in the inverse weak value regime so that $\phi < k\sigma \ll 1$.  We also expand to linear order in $kQ$, but make no assumption about the relative size of $kQ$ to the other parameters.  Finally, since $\ell_2$ is completely under the experimenter's control, and can be made arbitrarily small, we take $\ell_2=0$.  This yields a per-event Fisher information of 
\begin{align}
	\mathcal{I}(\phi) &\approx \frac{4k_0^2}{k^2} \cdot \frac{4k_0^2\sigma^2 + \ell_1^2\left(\frac{1}{\sigma^2} -4Q^2\right)}{\left(\frac{\ell_1^2}{\sigma^2} + 4k_0^2\sigma^2\right)^2}.
\end{align}
In the limit of $k_0\sigma \gg \ell_1/\sigma$ and $k_0\sigma \gg \ell_1Q$, this reduces to 
\begin{align}
	\mathcal{I}(\phi) &= \frac{1}{k^2\sigma^2},
\end{align}
which is simply the inverse of the postselection probability.  Hence, the total Fisher information for $\nu$ independent events becomes 
\begin{align}
	\mathcal{I}_\nu(\phi) &= p_F\nu \mathcal{I}(\phi), \nonumber \\
	&\approx \nu.
\end{align}
Amplification in the inverse weak value regime under the effect of angular jitter noise allows for recovery of the full noiseless Fisher information for large beam width $\sigma$, as in the case of additive Gaussian white noise considered in Sec.~\ref{sec:gaussian noise}.  This is confirmed numerically using the full state \eqref{eq:jitter state}, with results shown in Fig.~\ref{fig:jitter plot}.

\section{Frequency measurements in the inverse weak value regime}\label{sec:frequency}
It is straightforward to extend the above results for precision phase estimation to frequency estimation.  Here we treat two separate optical frequency measurement schemes which are robust to technical noise.

\subsection{Prism method}
We can convert the phase measurements above to a frequency shift measurement simply by replacing the SBC and HWP in Fig.~\ref{fig:jitter setup} with a prism.  If we take the optical axis to be the $z$-direction and the momentum shift from the tilted mirror in the $x$-direction, we choose the prism to cause a small frequency dependent momentum shift $k_p(\omega)$ in the $y$-direction.  We will abbreviate this as $k_p$ in what follows for the sake of notational simplicity.  If our detector only measures shifts in the $x$-direction, the effect of the prism will be essentially the same as introducing a frequency dependent phase.  

Before the detector, the two-dimensional distribution becomes
\begin{align}
	p(x,y|k_p) &= \frac{1}{p_f} \sin^2\left(kx + k_p y\right)\left|\ipr{x,y}{\varphi}\right|^2,
\end{align} 
where $\left|\ipr{x,y}{\varphi}\right|^2$ is an uncorrelated two-dimensional Gaussian distribution with zero mean and the same variance in both directions equal to $\sigma^2$ (\emph{i.e.,}~the covariance matrix for the initial state is proportional to the identity).  To first order in $k_p$ the postselection probability is unchanged from Eq.~\eqref{eq:postselection probability}.  If we are only measuring in the $x$-direction, the distribution becomes
\begin{align}
	\nonumber p(x|k_p) &= \int_{-\infty}^{\infty} \text{d}y~p(x,y|k_p), \\
	&\approx \frac{1}{p_f}\sin^2\left(kx +k_p \sigma \right)\left|\ipr{x}{\varphi}\right|^2.
\end{align}
Hence, the above analysis for phase measurements hold for this type of frequency measurement with the change $\phi \rightarrow 2k_p \sigma$, where the factor of two is simply due to our definitions of the relevant interaction Hamiltonians in Sec.~\ref{sec:inverse weak values}.  

The Fisher information per detected event for estimation of the parameter $k_p$ to lowest nonvanishing order in $k\sigma$ is equal to 
\begin{align}
	\mathcal{I}(k_p) &\approx \frac{2\sigma^2}{k^2\sigma^2}, 
\end{align} 
where the expression is no longer dimensionless since the parameter being estimated is not dimensionless.  The total Fisher information for $\nu$ events is then
\begin{align}
	\mathcal{I}_\nu(k_p) &= p_f \nu \mathcal{I}(k_p), \nonumber \\
	& \approx 2 \sigma^2 \nu .
\end{align}
To translate this to a minimum frequency resolution we note that 
\begin{align}
	k_p &\approx k_0 \theta, \nonumber \\
	& \approx k_0 \frac{\partial \theta}{\partial \omega}\Delta \omega,
\end{align}
where $k_0$ is the wavenumber, $\Delta \omega$ is the frequency shift relative to some reference frequency, and $\theta$ is the angle of deflection relative to the optical axis.  Since the estimator for $\Delta \omega$ differs from the one for $k_p$ by only a factor, it follows directly the Fisher information is given by 
\begin{align}
	\mathcal{I}_\nu(\Delta \omega) &= \left[k_0\frac{\partial \theta}{\partial \omega}\right]^{2}\mathcal{I}_\nu(k_p), \nonumber \\
	&= 2\left[k_0\sigma\frac{\partial \theta}{\partial \omega}\right]^{2}\nu.
\end{align}
Hence, the resolution of a precision frequency measurement can be made large for large $\sigma$ and for large $\partial\theta/\partial\omega$, as is the case for an atomic prism \cite{Starling2012}.

As an example, for a group velocity $v_g/c = 10^{-3}$ (achievable in an atomic prism \cite{Starling2012}), $\sigma = 1$ cm,  $\lambda = 780$ nm, a laser power of 1 mW and an integration time of 1 second, we have a minimum resolvable frequency shift $\Delta \omega \sim 1$ Hz.
\subsection{Group velocity delay method}
\begin{figure}
	\includegraphics[scale=.5]{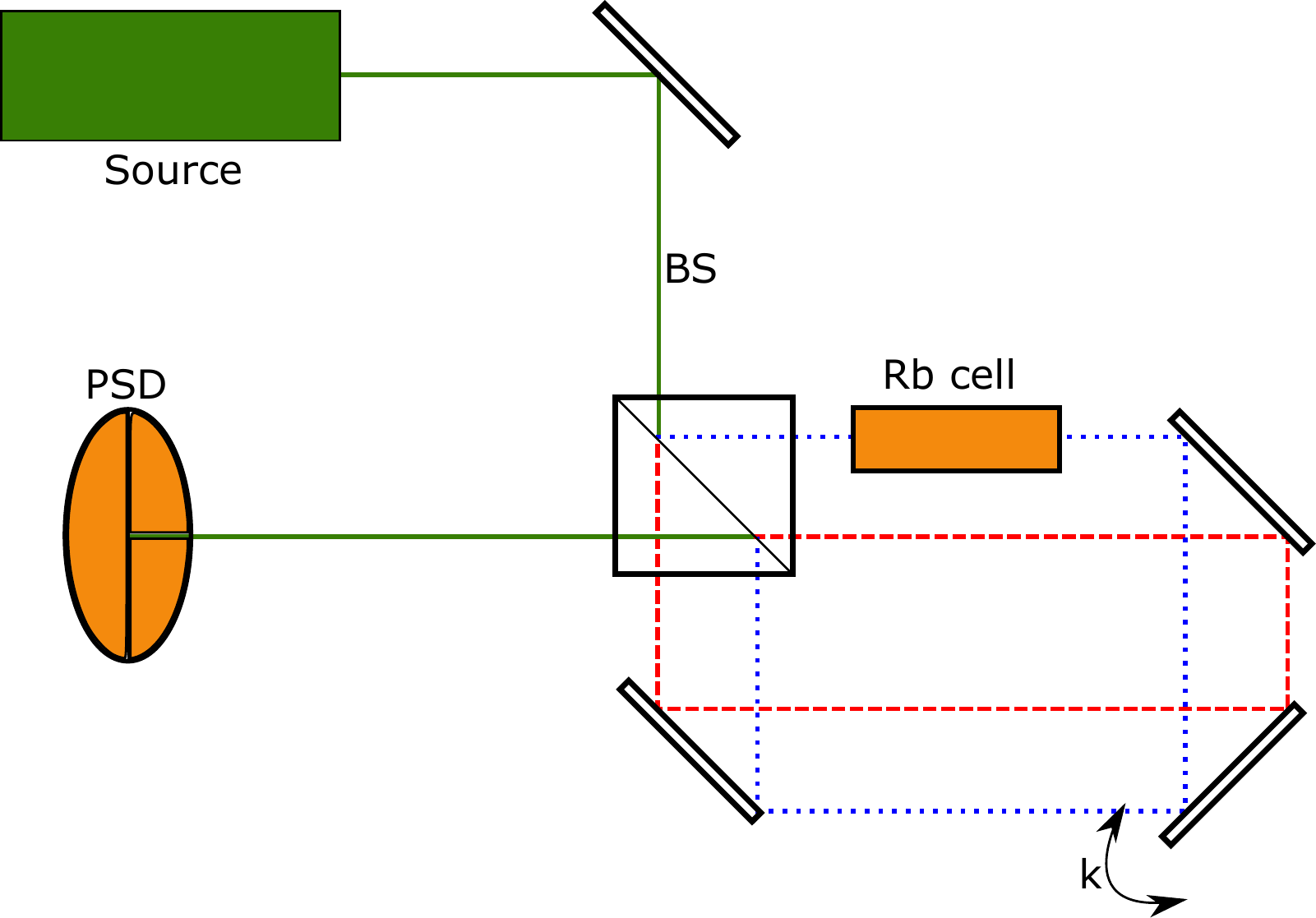}
\caption{An alternate geometry for precision frequency or phase measurements using a displaced Sagnac interferometer.  In the case of a frequency measurement a dispersive medium such as a warm rubidium (Rb) vapor must be used.  The blue dotted line represents the clockwise path in the interferometer and the red dashed line represents the counterclockwise path.}\label{fig:displaced sagnac}
\end{figure}
To generate a frequency dependent phase shift without introducing an unmeasured shift in the beam as above, we choose a slightly modified experimental geometry shown in Fig.~\ref{fig:displaced sagnac}.  The beam entering the interferometer is displaced from the symmetric axis so the counter-propagating paths do not overlap.  This allows us to place a dispersive medium in only a single path, which we again take to introduce a relative phase $\phi = (n(\omega)-1)k_0 d$ between the paths, where $n(\omega)$ is the frequency dependent index of refraction, $k_0$ is the beam wavenumber in vacuum, and $d$ is the path length through the medium.   

Since the measurement operators in Sec.~\ref{sec:inverse weak values} are unchanged up to meaningless global phase factors, we are able to write the probability distribution as 
\begin{align}
	p(x|\omega) &= \frac{1}{p_f}\sin^2\left(kx + \frac{\phi(\omega)}{2}\right)\left|\ipr{x}{\varphi}\right|^2.
\end{align}
The Fisher information per detected event for the parameter $\omega$ is then
\begin{align}
	\nonumber \mathcal{I}(\omega) &= \frac{1}{p_f}\left[\left(\frac{\partial \phi}{\partial \omega}\right)^2 - e^{-2k^2\sigma^2}\sin(\phi) \frac{\partial^2 \phi}{\partial \omega^2}\right], \\
	\nonumber &\approx \frac{1}{p_f}\left(\frac{\partial \phi}{\partial \omega}\right)^2, \\
	\nonumber &= \frac{d^2}{p_f}\left[\frac{1}{v_g} - \frac{1}{c}\right]^2, \\
	&\approx \frac{\tau_g^2}{p_f},
\end{align}
where $v_g \ll c$ is the group velocity in the dispersive medium, and $\tau_g$ represents the time it takes a wave envelope to travel a distance $d$ at the group velocity (the group delay).  

As above, for $\nu$ total input events, we multiply this expression by $p_f \nu$ to get the total Fisher information
\begin{align}
	\mathcal{I}_\nu(\omega) &= \nu\tau_g^2.
\end{align}
For a warm rubidium vapor, $v_g/c$ can potentially be on the order of $10^{-7}$ \cite{Boyd2001}.  Here we will use a readily achievable value of $10^{-3}$, as in the previous subsection.  With a cell width of 1 cm, $\lambda = 780$ nm, a laser power of 1 mW and an integration time of 1 second, we have a minimum resolvable frequency shift of $\Delta \omega \sim 10^{-1}$ Hz. 

\section{Conclusion}
As in the weak value regime, precision phase measurements in the inverse weak value regime are robust to several types of technical noise.  In particular, the type of measurements considered here can approach the full noiseless Fisher information under additive Gaussian technical noise and angular jitter noise if one is free to increase the variance $\sigma^2$ of the initial state arbitrarily.  Our results do not crucially rely on a specific experimental geometry, and will generalize well to a wide range of weak value based phase measurements. 

Additionally, we have shown that the same techniques used to estimate phase with weak measurements can also be used for frequency estimation, with the same robustness to technical noise.  This analysis demonstrates the usefulness of this kind of relative frequency metrology with a possible application to laser-locking.  Further research directions include the potential cooling of the mirrors to reduce thermal vibrations.

\section{Acknowledgments}
This work was supported by DRS Technologies and Army Research Office Grant No. W911NF-13-1-0402. 

\bibliography{inverseWV}

\begin{thebibliography}{39}%
\makeatletter
\providecommand \@ifxundefined [1]{%
 \@ifx{#1\undefined}
}%
\providecommand \@ifnum [1]{%
 \ifnum #1\expandafter \@firstoftwo
 \else \expandafter \@secondoftwo
 \fi
}%
\providecommand \@ifx [1]{%
 \ifx #1\expandafter \@firstoftwo
 \else \expandafter \@secondoftwo
 \fi
}%
\providecommand \natexlab [1]{#1}%
\providecommand \enquote  [1]{``#1''}%
\providecommand \bibnamefont  [1]{#1}%
\providecommand \bibfnamefont [1]{#1}%
\providecommand \citenamefont [1]{#1}%
\providecommand \href@noop [0]{\@secondoftwo}%
\providecommand \href [0]{\begingroup \@sanitize@url \@href}%
\providecommand \@href[1]{\@@startlink{#1}\@@href}%
\providecommand \@@href[1]{\endgroup#1\@@endlink}%
\providecommand \@sanitize@url [0]{\catcode `\\12\catcode `\$12\catcode
  `\&12\catcode `\#12\catcode `\^12\catcode `\_12\catcode `\%12\relax}%
\providecommand \@@startlink[1]{}%
\providecommand \@@endlink[0]{}%
\providecommand \url  [0]{\begingroup\@sanitize@url \@url }%
\providecommand \@url [1]{\endgroup\@href {#1}{\urlprefix }}%
\providecommand \urlprefix  [0]{URL }%
\providecommand \Eprint [0]{\href }%
\providecommand \doibase [0]{http://dx.doi.org/}%
\providecommand \selectlanguage [0]{\@gobble}%
\providecommand \bibinfo  [0]{\@secondoftwo}%
\providecommand \bibfield  [0]{\@secondoftwo}%
\providecommand \translation [1]{[#1]}%
\providecommand \BibitemOpen [0]{}%
\providecommand \bibitemStop [0]{}%
\providecommand \bibitemNoStop [0]{.\EOS\space}%
\providecommand \EOS [0]{\spacefactor3000\relax}%
\providecommand \BibitemShut  [1]{\csname bibitem#1\endcsname}%
\let\auto@bib@innerbib\@empty
\bibitem [{\citenamefont {Aharonov}\ \emph {et~al.}(1988)\citenamefont
  {Aharonov}, \citenamefont {Albert},\ and\ \citenamefont
  {Vaidman}}]{Aharonov1988}%
  \BibitemOpen
  \bibfield  {author} {\bibinfo {author} {\bibfnamefont {Y.}~\bibnamefont
  {Aharonov}}, \bibinfo {author} {\bibfnamefont {D.~Z.}\ \bibnamefont
  {Albert}}, \ and\ \bibinfo {author} {\bibfnamefont {L.}~\bibnamefont
  {Vaidman}},\ }\href {\doibase 10.1103/PhysRevLett.60.1351} {\bibfield
  {journal} {\bibinfo  {journal} {Phys. Rev. Lett.}\ }\textbf {\bibinfo
  {volume} {60}},\ \bibinfo {pages} {1351} (\bibinfo {year}
  {1988})}\BibitemShut {NoStop}%
\bibitem [{\citenamefont {Hosten}\ and\ \citenamefont
  {Kwiat}(2008)}]{Hosten2008}%
  \BibitemOpen
  \bibfield  {author} {\bibinfo {author} {\bibfnamefont {O.}~\bibnamefont
  {Hosten}}\ and\ \bibinfo {author} {\bibfnamefont {P.}~\bibnamefont {Kwiat}},\
  }\href@noop {} {\bibfield  {journal} {\bibinfo  {journal} {Science}\ }\textbf
  {\bibinfo {volume} {319}},\ \bibinfo {pages} {787} (\bibinfo {year}
  {2008})}\BibitemShut {NoStop}%
\bibitem [{\citenamefont {Dixon}\ \emph {et~al.}(2009)\citenamefont {Dixon},
  \citenamefont {Starling}, \citenamefont {Jordan},\ and\ \citenamefont
  {Howell}}]{Dixon2009}%
  \BibitemOpen
  \bibfield  {author} {\bibinfo {author} {\bibfnamefont {P.~B.}\ \bibnamefont
  {Dixon}}, \bibinfo {author} {\bibfnamefont {D.~J.}\ \bibnamefont {Starling}},
  \bibinfo {author} {\bibfnamefont {A.~N.}\ \bibnamefont {Jordan}}, \ and\
  \bibinfo {author} {\bibfnamefont {J.~C.}\ \bibnamefont {Howell}},\
  }\href@noop {} {\bibfield  {journal} {\bibinfo  {journal} {Phys. Rev. Lett.}\
  }\textbf {\bibinfo {volume} {102}},\ \bibinfo {pages} {173601} (\bibinfo
  {year} {2009})}\BibitemShut {NoStop}%
\bibitem [{\citenamefont {Starling}\ \emph {et~al.}(2009)\citenamefont
  {Starling}, \citenamefont {Dixon}, \citenamefont {Jordan},\ and\
  \citenamefont {Howell}}]{Starling2009}%
  \BibitemOpen
  \bibfield  {author} {\bibinfo {author} {\bibfnamefont {D.~J.}\ \bibnamefont
  {Starling}}, \bibinfo {author} {\bibfnamefont {P.~B.}\ \bibnamefont {Dixon}},
  \bibinfo {author} {\bibfnamefont {A.~N.}\ \bibnamefont {Jordan}}, \ and\
  \bibinfo {author} {\bibfnamefont {J.~C.}\ \bibnamefont {Howell}},\
  }\href@noop {} {\bibfield  {journal} {\bibinfo  {journal} {Phys. Rev. A}\
  }\textbf {\bibinfo {volume} {80}},\ \bibinfo {pages} {041803} (\bibinfo
  {year} {2009})}\BibitemShut {NoStop}%
\bibitem [{\citenamefont {Starling}\ \emph
  {et~al.}(2010{\natexlab{a}})\citenamefont {Starling}, \citenamefont {Dixon},
  \citenamefont {Jordan},\ and\ \citenamefont {Howell}}]{Starling2010}%
  \BibitemOpen
  \bibfield  {author} {\bibinfo {author} {\bibfnamefont {D.~J.}\ \bibnamefont
  {Starling}}, \bibinfo {author} {\bibfnamefont {P.~B.}\ \bibnamefont {Dixon}},
  \bibinfo {author} {\bibfnamefont {A.~N.}\ \bibnamefont {Jordan}}, \ and\
  \bibinfo {author} {\bibfnamefont {J.~C.}\ \bibnamefont {Howell}},\
  }\href@noop {} {\bibfield  {journal} {\bibinfo  {journal} {Phys. Rev. A}\
  }\textbf {\bibinfo {volume} {82}},\ \bibinfo {pages} {063822} (\bibinfo
  {year} {2010}{\natexlab{a}})}\BibitemShut {NoStop}%
\bibitem [{\citenamefont {Starling}\ \emph
  {et~al.}(2010{\natexlab{b}})\citenamefont {Starling}, \citenamefont {Dixon},
  \citenamefont {Williams}, \citenamefont {Jordan},\ and\ \citenamefont
  {Howell}}]{Starling2010a}%
  \BibitemOpen
  \bibfield  {author} {\bibinfo {author} {\bibfnamefont {D.~J.}\ \bibnamefont
  {Starling}}, \bibinfo {author} {\bibfnamefont {P.~B.}\ \bibnamefont {Dixon}},
  \bibinfo {author} {\bibfnamefont {N.~S.}\ \bibnamefont {Williams}}, \bibinfo
  {author} {\bibfnamefont {A.~N.}\ \bibnamefont {Jordan}}, \ and\ \bibinfo
  {author} {\bibfnamefont {J.~C.}\ \bibnamefont {Howell}},\ }\href@noop {}
  {\bibfield  {journal} {\bibinfo  {journal} {Phys. Rev. A}\ }\textbf {\bibinfo
  {volume} {82}},\ \bibinfo {pages} {011802(R)} (\bibinfo {year}
  {2010}{\natexlab{b}})}\BibitemShut {NoStop}%
\bibitem [{\citenamefont {Howell}\ \emph {et~al.}(2010)\citenamefont {Howell},
  \citenamefont {Starling}, \citenamefont {Dixon}, \citenamefont {Vudyasetu},\
  and\ \citenamefont {Jordan}}]{Howell2010}%
  \BibitemOpen
  \bibfield  {author} {\bibinfo {author} {\bibfnamefont {J.~C.}\ \bibnamefont
  {Howell}}, \bibinfo {author} {\bibfnamefont {D.~J.}\ \bibnamefont
  {Starling}}, \bibinfo {author} {\bibfnamefont {P.~B.}\ \bibnamefont {Dixon}},
  \bibinfo {author} {\bibfnamefont {P.~K.}\ \bibnamefont {Vudyasetu}}, \ and\
  \bibinfo {author} {\bibfnamefont {A.~N.}\ \bibnamefont {Jordan}},\
  }\href@noop {} {\bibfield  {journal} {\bibinfo  {journal} {Phys. Rev. A}\
  }\textbf {\bibinfo {volume} {81}},\ \bibinfo {pages} {033813} (\bibinfo
  {year} {2010})}\BibitemShut {NoStop}%
\bibitem [{\citenamefont {Hogan}\ \emph {et~al.}(2011)\citenamefont {Hogan},
  \citenamefont {Hammer}, \citenamefont {Chiow}, \citenamefont {Dickerson},
  \citenamefont {Johnson}, \citenamefont {Kovachy}, \citenamefont
  {Sugarbaker},\ and\ \citenamefont {Kasevich}}]{Hogan:11}%
  \BibitemOpen
  \bibfield  {author} {\bibinfo {author} {\bibfnamefont {J.~M.}\ \bibnamefont
  {Hogan}}, \bibinfo {author} {\bibfnamefont {J.}~\bibnamefont {Hammer}},
  \bibinfo {author} {\bibfnamefont {S.-W.}\ \bibnamefont {Chiow}}, \bibinfo
  {author} {\bibfnamefont {S.}~\bibnamefont {Dickerson}}, \bibinfo {author}
  {\bibfnamefont {D.~M.~S.}\ \bibnamefont {Johnson}}, \bibinfo {author}
  {\bibfnamefont {T.}~\bibnamefont {Kovachy}}, \bibinfo {author} {\bibfnamefont
  {A.}~\bibnamefont {Sugarbaker}}, \ and\ \bibinfo {author} {\bibfnamefont
  {M.~A.}\ \bibnamefont {Kasevich}},\ }\href {\doibase 10.1364/OL.36.001698}
  {\bibfield  {journal} {\bibinfo  {journal} {Opt. Lett.}\ }\textbf {\bibinfo
  {volume} {36}},\ \bibinfo {pages} {1698} (\bibinfo {year}
  {2011})}\BibitemShut {NoStop}%
\bibitem [{\citenamefont {Pfeifer}\ and\ \citenamefont
  {Fischer}(2011)}]{Pfeifer:11}%
  \BibitemOpen
  \bibfield  {author} {\bibinfo {author} {\bibfnamefont {M.}~\bibnamefont
  {Pfeifer}}\ and\ \bibinfo {author} {\bibfnamefont {P.}~\bibnamefont
  {Fischer}},\ }\href {\doibase 10.1364/OE.19.016508} {\bibfield  {journal}
  {\bibinfo  {journal} {Opt. Express}\ }\textbf {\bibinfo {volume} {19}},\
  \bibinfo {pages} {16508} (\bibinfo {year} {2011})}\BibitemShut {NoStop}%
\bibitem [{\citenamefont {Egan}\ and\ \citenamefont {Stone}(2012)}]{Egan:12}%
  \BibitemOpen
  \bibfield  {author} {\bibinfo {author} {\bibfnamefont {P.}~\bibnamefont
  {Egan}}\ and\ \bibinfo {author} {\bibfnamefont {J.~A.}\ \bibnamefont
  {Stone}},\ }\href {\doibase 10.1364/OL.37.004991} {\bibfield  {journal}
  {\bibinfo  {journal} {Opt. Lett.}\ }\textbf {\bibinfo {volume} {37}},\
  \bibinfo {pages} {4991} (\bibinfo {year} {2012})}\BibitemShut {NoStop}%
\bibitem [{\citenamefont {Gorodetski}\ \emph {et~al.}(2012)\citenamefont
  {Gorodetski}, \citenamefont {Bliokh}, \citenamefont {Stein}, \citenamefont
  {Genet}, \citenamefont {Shitrit}, \citenamefont {Kleiner}, \citenamefont
  {Hasman},\ and\ \citenamefont {Ebbesen}}]{PhysRevLett.109.013901}%
  \BibitemOpen
  \bibfield  {author} {\bibinfo {author} {\bibfnamefont {Y.}~\bibnamefont
  {Gorodetski}}, \bibinfo {author} {\bibfnamefont {K.~Y.}\ \bibnamefont
  {Bliokh}}, \bibinfo {author} {\bibfnamefont {B.}~\bibnamefont {Stein}},
  \bibinfo {author} {\bibfnamefont {C.}~\bibnamefont {Genet}}, \bibinfo
  {author} {\bibfnamefont {N.}~\bibnamefont {Shitrit}}, \bibinfo {author}
  {\bibfnamefont {V.}~\bibnamefont {Kleiner}}, \bibinfo {author} {\bibfnamefont
  {E.}~\bibnamefont {Hasman}}, \ and\ \bibinfo {author} {\bibfnamefont {T.~W.}\
  \bibnamefont {Ebbesen}},\ }\href {\doibase 10.1103/PhysRevLett.109.013901}
  {\bibfield  {journal} {\bibinfo  {journal} {Phys. Rev. Lett.}\ }\textbf
  {\bibinfo {volume} {109}},\ \bibinfo {pages} {013901} (\bibinfo {year}
  {2012})}\BibitemShut {NoStop}%
\bibitem [{\citenamefont {Zhou}\ \emph {et~al.}(2012)\citenamefont {Zhou},
  \citenamefont {Xiao}, \citenamefont {Luo},\ and\ \citenamefont
  {Wen}}]{Zhou2012}%
  \BibitemOpen
  \bibfield  {author} {\bibinfo {author} {\bibfnamefont {X.}~\bibnamefont
  {Zhou}}, \bibinfo {author} {\bibfnamefont {Z.}~\bibnamefont {Xiao}}, \bibinfo
  {author} {\bibfnamefont {H.}~\bibnamefont {Luo}}, \ and\ \bibinfo {author}
  {\bibfnamefont {S.}~\bibnamefont {Wen}},\ }\href {\doibase
  10.1103/PhysRevA.85.043809} {\bibfield  {journal} {\bibinfo  {journal} {Phys.
  Rev. A}\ }\textbf {\bibinfo {volume} {85}},\ \bibinfo {pages} {043809}
  (\bibinfo {year} {2012})}\BibitemShut {NoStop}%
\bibitem [{\citenamefont {Str\"ubi}\ and\ \citenamefont
  {Bruder}(2013)}]{Strubi2013}%
  \BibitemOpen
  \bibfield  {author} {\bibinfo {author} {\bibfnamefont {G.}~\bibnamefont
  {Str\"ubi}}\ and\ \bibinfo {author} {\bibfnamefont {C.}~\bibnamefont
  {Bruder}},\ }\href {\doibase 10.1103/PhysRevLett.110.083605} {\bibfield
  {journal} {\bibinfo  {journal} {Phys. Rev. Lett.}\ }\textbf {\bibinfo
  {volume} {110}},\ \bibinfo {pages} {083605} (\bibinfo {year}
  {2013})}\BibitemShut {NoStop}%
\bibitem [{\citenamefont {Viza}\ \emph {et~al.}(2013)\citenamefont {Viza},
  \citenamefont {Mart\'{i}nez-Rinc\'{o}n}, \citenamefont {Howland},
  \citenamefont {Frostig}, \citenamefont {Shomroni}, \citenamefont {Dayan},\
  and\ \citenamefont {Howell}}]{Viza:13}%
  \BibitemOpen
  \bibfield  {author} {\bibinfo {author} {\bibfnamefont {G.~I.}\ \bibnamefont
  {Viza}}, \bibinfo {author} {\bibfnamefont {J.}~\bibnamefont
  {Mart\'{i}nez-Rinc\'{o}n}}, \bibinfo {author} {\bibfnamefont {G.~A.}\
  \bibnamefont {Howland}}, \bibinfo {author} {\bibfnamefont {H.}~\bibnamefont
  {Frostig}}, \bibinfo {author} {\bibfnamefont {I.}~\bibnamefont {Shomroni}},
  \bibinfo {author} {\bibfnamefont {B.}~\bibnamefont {Dayan}}, \ and\ \bibinfo
  {author} {\bibfnamefont {J.~C.}\ \bibnamefont {Howell}},\ }\href {\doibase
  10.1364/OL.38.002949} {\bibfield  {journal} {\bibinfo  {journal} {Opt.
  Lett.}\ }\textbf {\bibinfo {volume} {38}},\ \bibinfo {pages} {2949} (\bibinfo
  {year} {2013})}\BibitemShut {NoStop}%
\bibitem [{\citenamefont {Xu}\ \emph {et~al.}(2013)\citenamefont {Xu},
  \citenamefont {Kedem}, \citenamefont {Sun}, \citenamefont {Vaidman},
  \citenamefont {Li},\ and\ \citenamefont {Guo}}]{PhysRevLettv111p033604}%
  \BibitemOpen
  \bibfield  {author} {\bibinfo {author} {\bibfnamefont {X.-Y.}\ \bibnamefont
  {Xu}}, \bibinfo {author} {\bibfnamefont {Y.}~\bibnamefont {Kedem}}, \bibinfo
  {author} {\bibfnamefont {K.}~\bibnamefont {Sun}}, \bibinfo {author}
  {\bibfnamefont {L.}~\bibnamefont {Vaidman}}, \bibinfo {author} {\bibfnamefont
  {C.-F.}\ \bibnamefont {Li}}, \ and\ \bibinfo {author} {\bibfnamefont {G.-C.}\
  \bibnamefont {Guo}},\ }\href {\doibase 10.1103/PhysRevLett.111.033604}
  {\bibfield  {journal} {\bibinfo  {journal} {Phys. Rev. Lett.}\ }\textbf
  {\bibinfo {volume} {111}},\ \bibinfo {pages} {033604} (\bibinfo {year}
  {2013})}\BibitemShut {NoStop}%
\bibitem [{\citenamefont {Zhou}\ \emph {et~al.}(2013)\citenamefont {Zhou},
  \citenamefont {Turek}, \citenamefont {Sun},\ and\ \citenamefont
  {Nori}}]{Zhou2013}%
  \BibitemOpen
  \bibfield  {author} {\bibinfo {author} {\bibfnamefont {L.}~\bibnamefont
  {Zhou}}, \bibinfo {author} {\bibfnamefont {Y.}~\bibnamefont {Turek}},
  \bibinfo {author} {\bibfnamefont {C.~P.}\ \bibnamefont {Sun}}, \ and\
  \bibinfo {author} {\bibfnamefont {F.}~\bibnamefont {Nori}},\ }\href {\doibase
  10.1103/PhysRevA.88.053815} {\bibfield  {journal} {\bibinfo  {journal} {Phys.
  Rev. A}\ }\textbf {\bibinfo {volume} {88}},\ \bibinfo {pages} {053815}
  (\bibinfo {year} {2013})}\BibitemShut {NoStop}%
\bibitem [{\citenamefont {Maga\~na Loaiza}\ \emph {et~al.}(2014)\citenamefont
  {Maga\~na Loaiza}, \citenamefont {Mirhosseini}, \citenamefont {Rodenburg},\
  and\ \citenamefont {Boyd}}]{PhysRevLett.112.200401}%
  \BibitemOpen
  \bibfield  {author} {\bibinfo {author} {\bibfnamefont {O.~S.}\ \bibnamefont
  {Maga\~na Loaiza}}, \bibinfo {author} {\bibfnamefont {M.}~\bibnamefont
  {Mirhosseini}}, \bibinfo {author} {\bibfnamefont {B.}~\bibnamefont
  {Rodenburg}}, \ and\ \bibinfo {author} {\bibfnamefont {R.~W.}\ \bibnamefont
  {Boyd}},\ }\href {\doibase 10.1103/PhysRevLett.112.200401} {\bibfield
  {journal} {\bibinfo  {journal} {Phys. Rev. Lett.}\ }\textbf {\bibinfo
  {volume} {112}},\ \bibinfo {pages} {200401} (\bibinfo {year}
  {2014})}\BibitemShut {NoStop}%
\bibitem [{\citenamefont {Salazar-Serrano}\ \emph {et~al.}(2014)\citenamefont
  {Salazar-Serrano}, \citenamefont {Janner}, \citenamefont {Brunner},
  \citenamefont {Pruneri},\ and\ \citenamefont {Torres}}]{PhysRevA.89.012126}%
  \BibitemOpen
  \bibfield  {author} {\bibinfo {author} {\bibfnamefont {L.~J.}\ \bibnamefont
  {Salazar-Serrano}}, \bibinfo {author} {\bibfnamefont {D.}~\bibnamefont
  {Janner}}, \bibinfo {author} {\bibfnamefont {N.}~\bibnamefont {Brunner}},
  \bibinfo {author} {\bibfnamefont {V.}~\bibnamefont {Pruneri}}, \ and\
  \bibinfo {author} {\bibfnamefont {J.~P.}\ \bibnamefont {Torres}},\ }\href
  {\doibase 10.1103/PhysRevA.89.012126} {\bibfield  {journal} {\bibinfo
  {journal} {Phys. Rev. A}\ }\textbf {\bibinfo {volume} {89}},\ \bibinfo
  {pages} {012126} (\bibinfo {year} {2014})}\BibitemShut {NoStop}%
\bibitem [{\citenamefont {Salazar-Serrano}\ \emph {et~al.}(2015)\citenamefont
  {Salazar-Serrano}, \citenamefont {Barrera}, \citenamefont {Amaya},
  \citenamefont {Sales}, \citenamefont {Pruneri}, \citenamefont {Capmany},\
  and\ \citenamefont {Torres}}]{salazar2015enhancement}%
  \BibitemOpen
  \bibfield  {author} {\bibinfo {author} {\bibfnamefont {L.}~\bibnamefont
  {Salazar-Serrano}}, \bibinfo {author} {\bibfnamefont {D.}~\bibnamefont
  {Barrera}}, \bibinfo {author} {\bibfnamefont {W.}~\bibnamefont {Amaya}},
  \bibinfo {author} {\bibfnamefont {S.}~\bibnamefont {Sales}}, \bibinfo
  {author} {\bibfnamefont {V.}~\bibnamefont {Pruneri}}, \bibinfo {author}
  {\bibfnamefont {J.}~\bibnamefont {Capmany}}, \ and\ \bibinfo {author}
  {\bibfnamefont {J.}~\bibnamefont {Torres}},\ }\href@noop {} {\bibfield
  {journal} {\bibinfo  {journal} {Optics letters}\ }\textbf {\bibinfo {volume}
  {40}},\ \bibinfo {pages} {3962} (\bibinfo {year} {2015})}\BibitemShut
  {NoStop}%
\bibitem [{\citenamefont {Mart\'{i}nez-Rinc\'{o}n}\ \emph
  {et~al.}(2017)\citenamefont {Mart\'{i}nez-Rinc\'{o}n}, \citenamefont
  {Mullarkey}, \citenamefont {Viza}, \citenamefont {Liu},\ and\ \citenamefont
  {Howell}}]{Martinez-Rincon:17}%
  \BibitemOpen
  \bibfield  {author} {\bibinfo {author} {\bibfnamefont {J.}~\bibnamefont
  {Mart\'{i}nez-Rinc\'{o}n}}, \bibinfo {author} {\bibfnamefont {C.~A.}\
  \bibnamefont {Mullarkey}}, \bibinfo {author} {\bibfnamefont {G.~I.}\
  \bibnamefont {Viza}}, \bibinfo {author} {\bibfnamefont {W.-T.}\ \bibnamefont
  {Liu}}, \ and\ \bibinfo {author} {\bibfnamefont {J.~C.}\ \bibnamefont
  {Howell}},\ }\href {\doibase 10.1364/OL.42.002479} {\bibfield  {journal}
  {\bibinfo  {journal} {Opt. Lett.}\ }\textbf {\bibinfo {volume} {42}},\
  \bibinfo {pages} {2479} (\bibinfo {year} {2017})}\BibitemShut {NoStop}%
\bibitem [{\citenamefont {Dressel}\ \emph {et~al.}(2014)\citenamefont
  {Dressel}, \citenamefont {Malik}, \citenamefont {Miatto}, \citenamefont
  {Jordan},\ and\ \citenamefont {Boyd}}]{Dressel2014}%
  \BibitemOpen
  \bibfield  {author} {\bibinfo {author} {\bibfnamefont {J.}~\bibnamefont
  {Dressel}}, \bibinfo {author} {\bibfnamefont {M.}~\bibnamefont {Malik}},
  \bibinfo {author} {\bibfnamefont {F.~M.}\ \bibnamefont {Miatto}}, \bibinfo
  {author} {\bibfnamefont {A.~N.}\ \bibnamefont {Jordan}}, \ and\ \bibinfo
  {author} {\bibfnamefont {R.~W.}\ \bibnamefont {Boyd}},\ }\href {\doibase
  10.1103/RevModPhys.86.307} {\bibfield  {journal} {\bibinfo  {journal} {Rev.
  Mod. Phys.}\ }\textbf {\bibinfo {volume} {86}},\ \bibinfo {pages} {307}
  (\bibinfo {year} {2014})}\BibitemShut {NoStop}%
\bibitem [{\citenamefont {Pang}\ \emph {et~al.}(2014)\citenamefont {Pang},
  \citenamefont {Dressel},\ and\ \citenamefont {Brun}}]{Pang2014}%
  \BibitemOpen
  \bibfield  {author} {\bibinfo {author} {\bibfnamefont {S.}~\bibnamefont
  {Pang}}, \bibinfo {author} {\bibfnamefont {J.}~\bibnamefont {Dressel}}, \
  and\ \bibinfo {author} {\bibfnamefont {T.~A.}\ \bibnamefont {Brun}},\ }\href
  {\doibase 10.1103/PhysRevLett.113.030401} {\bibfield  {journal} {\bibinfo
  {journal} {Phys. Rev. Lett.}\ }\textbf {\bibinfo {volume} {113}},\ \bibinfo
  {pages} {030401} (\bibinfo {year} {2014})}\BibitemShut {NoStop}%
\bibitem [{\citenamefont {Pang}\ and\ \citenamefont {Brun}(2015)}]{Pang2015}%
  \BibitemOpen
  \bibfield  {author} {\bibinfo {author} {\bibfnamefont {S.}~\bibnamefont
  {Pang}}\ and\ \bibinfo {author} {\bibfnamefont {T.~A.}\ \bibnamefont
  {Brun}},\ }\href {\doibase 10.1103/PhysRevLett.115.120401} {\bibfield
  {journal} {\bibinfo  {journal} {Phys. Rev. Lett.}\ }\textbf {\bibinfo
  {volume} {115}},\ \bibinfo {pages} {120401} (\bibinfo {year}
  {2015})}\BibitemShut {NoStop}%
\bibitem [{\citenamefont {Jordan}\ \emph {et~al.}(2014)\citenamefont {Jordan},
  \citenamefont {Mart\'{\i}nez-Rinc\'on},\ and\ \citenamefont
  {Howell}}]{Jordan2014}%
  \BibitemOpen
  \bibfield  {author} {\bibinfo {author} {\bibfnamefont {A.~N.}\ \bibnamefont
  {Jordan}}, \bibinfo {author} {\bibfnamefont {J.}~\bibnamefont
  {Mart\'{\i}nez-Rinc\'on}}, \ and\ \bibinfo {author} {\bibfnamefont {J.~C.}\
  \bibnamefont {Howell}},\ }\href {\doibase 10.1103/PhysRevX.4.011031}
  {\bibfield  {journal} {\bibinfo  {journal} {Phys. Rev. X}\ }\textbf {\bibinfo
  {volume} {4}},\ \bibinfo {pages} {011031} (\bibinfo {year}
  {2014})}\BibitemShut {NoStop}%
\bibitem [{\citenamefont {Knee}\ and\ \citenamefont {Gauger}(2014)}]{Knee2014}%
  \BibitemOpen
  \bibfield  {author} {\bibinfo {author} {\bibfnamefont {G.~C.}\ \bibnamefont
  {Knee}}\ and\ \bibinfo {author} {\bibfnamefont {E.~M.}\ \bibnamefont
  {Gauger}},\ }\href {\doibase 10.1103/PhysRevX.4.011032} {\bibfield  {journal}
  {\bibinfo  {journal} {Phys. Rev. X}\ }\textbf {\bibinfo {volume} {4}},\
  \bibinfo {pages} {011032} (\bibinfo {year} {2014})}\BibitemShut {NoStop}%
\bibitem [{\citenamefont {Viza}\ \emph {et~al.}(2015)\citenamefont {Viza},
  \citenamefont {Mart{\'\i}nez-Rinc{\'o}n}, \citenamefont {Alves},
  \citenamefont {Jordan},\ and\ \citenamefont
  {Howell}}]{viza2015experimentally}%
  \BibitemOpen
  \bibfield  {author} {\bibinfo {author} {\bibfnamefont {G.~I.}\ \bibnamefont
  {Viza}}, \bibinfo {author} {\bibfnamefont {J.}~\bibnamefont
  {Mart{\'\i}nez-Rinc{\'o}n}}, \bibinfo {author} {\bibfnamefont {G.~B.}\
  \bibnamefont {Alves}}, \bibinfo {author} {\bibfnamefont {A.~N.}\ \bibnamefont
  {Jordan}}, \ and\ \bibinfo {author} {\bibfnamefont {J.~C.}\ \bibnamefont
  {Howell}},\ }\href@noop {} {\bibfield  {journal} {\bibinfo  {journal}
  {Physical Review A}\ }\textbf {\bibinfo {volume} {92}},\ \bibinfo {pages}
  {032127} (\bibinfo {year} {2015})}\BibitemShut {NoStop}%
\bibitem [{\citenamefont {Alves}\ \emph {et~al.}(2015)\citenamefont {Alves},
  \citenamefont {Escher}, \citenamefont {de~Matos~Filho}, \citenamefont
  {Zagury},\ and\ \citenamefont {Davidovich}}]{PhysRevA.91.062107}%
  \BibitemOpen
  \bibfield  {author} {\bibinfo {author} {\bibfnamefont {G.~B.}\ \bibnamefont
  {Alves}}, \bibinfo {author} {\bibfnamefont {B.~M.}\ \bibnamefont {Escher}},
  \bibinfo {author} {\bibfnamefont {R.~L.}\ \bibnamefont {de~Matos~Filho}},
  \bibinfo {author} {\bibfnamefont {N.}~\bibnamefont {Zagury}}, \ and\ \bibinfo
  {author} {\bibfnamefont {L.}~\bibnamefont {Davidovich}},\ }\href {\doibase
  10.1103/PhysRevA.91.062107} {\bibfield  {journal} {\bibinfo  {journal} {Phys.
  Rev. A}\ }\textbf {\bibinfo {volume} {91}},\ \bibinfo {pages} {062107}
  (\bibinfo {year} {2015})}\BibitemShut {NoStop}%
\bibitem [{\citenamefont {Torres}\ and\ \citenamefont
  {Salazar-Serrano}(2016)}]{torres2016weak}%
  \BibitemOpen
  \bibfield  {author} {\bibinfo {author} {\bibfnamefont {J.~P.}\ \bibnamefont
  {Torres}}\ and\ \bibinfo {author} {\bibfnamefont {L.~J.}\ \bibnamefont
  {Salazar-Serrano}},\ }\href@noop {} {\bibfield  {journal} {\bibinfo
  {journal} {Scientific reports}\ }\textbf {\bibinfo {volume} {6}},\ \bibinfo
  {pages} {19702} (\bibinfo {year} {2016})}\BibitemShut {NoStop}%
\bibitem [{\citenamefont {Harris}\ \emph {et~al.}(2016)\citenamefont {Harris},
  \citenamefont {Boyd},\ and\ \citenamefont {Lundeen}}]{harris2016weak}%
  \BibitemOpen
  \bibfield  {author} {\bibinfo {author} {\bibfnamefont {J.}~\bibnamefont
  {Harris}}, \bibinfo {author} {\bibfnamefont {R.~W.}\ \bibnamefont {Boyd}}, \
  and\ \bibinfo {author} {\bibfnamefont {J.~S.}\ \bibnamefont {Lundeen}},\
  }\href@noop {} {\bibfield  {journal} {\bibinfo  {journal} {arXiv preprint
  arXiv:1612.04327}\ } (\bibinfo {year} {2016})}\BibitemShut {NoStop}%
\bibitem [{\citenamefont {Pang}\ \emph {et~al.}(2016)\citenamefont {Pang},
  \citenamefont {Alonso}, \citenamefont {Brun},\ and\ \citenamefont
  {Jordan}}]{Pang2016}%
  \BibitemOpen
  \bibfield  {author} {\bibinfo {author} {\bibfnamefont {S.}~\bibnamefont
  {Pang}}, \bibinfo {author} {\bibfnamefont {J.~R.~G.}\ \bibnamefont {Alonso}},
  \bibinfo {author} {\bibfnamefont {T.~A.}\ \bibnamefont {Brun}}, \ and\
  \bibinfo {author} {\bibfnamefont {A.~N.}\ \bibnamefont {Jordan}},\ }\href
  {\doibase 10.1103/PhysRevA.94.012329} {\bibfield  {journal} {\bibinfo
  {journal} {Phys. Rev. A}\ }\textbf {\bibinfo {volume} {94}},\ \bibinfo
  {pages} {012329} (\bibinfo {year} {2016})}\BibitemShut {NoStop}%
\bibitem [{\citenamefont {Dressel}\ \emph
  {et~al.}(2013{\natexlab{a}})\citenamefont {Dressel}, \citenamefont {Lyons},
  \citenamefont {Jordan}, \citenamefont {Graham},\ and\ \citenamefont
  {Kwiat}}]{dressel2013strengthening}%
  \BibitemOpen
  \bibfield  {author} {\bibinfo {author} {\bibfnamefont {J.}~\bibnamefont
  {Dressel}}, \bibinfo {author} {\bibfnamefont {K.}~\bibnamefont {Lyons}},
  \bibinfo {author} {\bibfnamefont {A.~N.}\ \bibnamefont {Jordan}}, \bibinfo
  {author} {\bibfnamefont {T.~M.}\ \bibnamefont {Graham}}, \ and\ \bibinfo
  {author} {\bibfnamefont {P.~G.}\ \bibnamefont {Kwiat}},\ }\href@noop {}
  {\bibfield  {journal} {\bibinfo  {journal} {Physical Review A}\ }\textbf
  {\bibinfo {volume} {88}},\ \bibinfo {pages} {023821} (\bibinfo {year}
  {2013}{\natexlab{a}})}\BibitemShut {NoStop}%
\bibitem [{\citenamefont {Lyons}\ \emph
  {et~al.}(2015{\natexlab{a}})\citenamefont {Lyons}, \citenamefont {Dressel},
  \citenamefont {Jordan}, \citenamefont {Howell},\ and\ \citenamefont
  {Kwiat}}]{lyons2015power}%
  \BibitemOpen
  \bibfield  {author} {\bibinfo {author} {\bibfnamefont {K.}~\bibnamefont
  {Lyons}}, \bibinfo {author} {\bibfnamefont {J.}~\bibnamefont {Dressel}},
  \bibinfo {author} {\bibfnamefont {A.~N.}\ \bibnamefont {Jordan}}, \bibinfo
  {author} {\bibfnamefont {J.~C.}\ \bibnamefont {Howell}}, \ and\ \bibinfo
  {author} {\bibfnamefont {P.~G.}\ \bibnamefont {Kwiat}},\ }\href@noop {}
  {\bibfield  {journal} {\bibinfo  {journal} {Physical review letters}\
  }\textbf {\bibinfo {volume} {114}},\ \bibinfo {pages} {170801} (\bibinfo
  {year} {2015}{\natexlab{a}})}\BibitemShut {NoStop}%
\bibitem [{\citenamefont {Byard}\ \emph {et~al.}(2015)\citenamefont {Byard},
  \citenamefont {Graham}, \citenamefont {Jordan},\ and\ \citenamefont
  {Kwiat}}]{byard2015pulse}%
  \BibitemOpen
  \bibfield  {author} {\bibinfo {author} {\bibfnamefont {C.}~\bibnamefont
  {Byard}}, \bibinfo {author} {\bibfnamefont {T.}~\bibnamefont {Graham}},
  \bibinfo {author} {\bibfnamefont {A.}~\bibnamefont {Jordan}}, \ and\ \bibinfo
  {author} {\bibfnamefont {P.~G.}\ \bibnamefont {Kwiat}},\ }in\ \href@noop {}
  {\emph {\bibinfo {booktitle} {Frontiers in Optics}}}\ (\bibinfo
  {organization} {Optical Society of America},\ \bibinfo {year} {2015})\ pp.\
  \bibinfo {pages} {JW2A--69}\BibitemShut {NoStop}%
\bibitem [{\citenamefont {Wang}\ \emph {et~al.}(2016)\citenamefont {Wang},
  \citenamefont {Tang}, \citenamefont {Hu}, \citenamefont {Wang}, \citenamefont
  {Yu}, \citenamefont {Zhou}, \citenamefont {Cheng}, \citenamefont {Xu},
  \citenamefont {Fang}, \citenamefont {Wu} \emph
  {et~al.}}]{wang2016experimental}%
  \BibitemOpen
  \bibfield  {author} {\bibinfo {author} {\bibfnamefont {Y.-T.}\ \bibnamefont
  {Wang}}, \bibinfo {author} {\bibfnamefont {J.-S.}\ \bibnamefont {Tang}},
  \bibinfo {author} {\bibfnamefont {G.}~\bibnamefont {Hu}}, \bibinfo {author}
  {\bibfnamefont {J.}~\bibnamefont {Wang}}, \bibinfo {author} {\bibfnamefont
  {S.}~\bibnamefont {Yu}}, \bibinfo {author} {\bibfnamefont {Z.-Q.}\
  \bibnamefont {Zhou}}, \bibinfo {author} {\bibfnamefont {Z.-D.}\ \bibnamefont
  {Cheng}}, \bibinfo {author} {\bibfnamefont {J.-S.}\ \bibnamefont {Xu}},
  \bibinfo {author} {\bibfnamefont {S.-Z.}\ \bibnamefont {Fang}}, \bibinfo
  {author} {\bibfnamefont {Q.-L.}\ \bibnamefont {Wu}},  \emph {et~al.},\
  }\href@noop {} {\bibfield  {journal} {\bibinfo  {journal} {Physical Review
  Letters}\ }\textbf {\bibinfo {volume} {117}},\ \bibinfo {pages} {230801}
  (\bibinfo {year} {2016})}\BibitemShut {NoStop}%
\bibitem [{\citenamefont {Dressel}\ \emph
  {et~al.}(2013{\natexlab{b}})\citenamefont {Dressel}, \citenamefont {Lyons},
  \citenamefont {Jordan}, \citenamefont {Graham},\ and\ \citenamefont
  {Kwiat}}]{Dressel2013}%
  \BibitemOpen
  \bibfield  {author} {\bibinfo {author} {\bibfnamefont {J.}~\bibnamefont
  {Dressel}}, \bibinfo {author} {\bibfnamefont {K.}~\bibnamefont {Lyons}},
  \bibinfo {author} {\bibfnamefont {A.~N.}\ \bibnamefont {Jordan}}, \bibinfo
  {author} {\bibfnamefont {T.}~\bibnamefont {Graham}}, \ and\ \bibinfo {author}
  {\bibfnamefont {P.}~\bibnamefont {Kwiat}},\ }\href@noop {} {\bibfield
  {journal} {\bibinfo  {journal} {Phys. Rev. A}\ }\textbf {\bibinfo {volume}
  {88}},\ \bibinfo {pages} {023821} (\bibinfo {year}
  {2013}{\natexlab{b}})}\BibitemShut {NoStop}%
\bibitem [{\citenamefont {Lyons}\ \emph
  {et~al.}(2015{\natexlab{b}})\citenamefont {Lyons}, \citenamefont {Dressel},
  \citenamefont {Jordan}, \citenamefont {Howell},\ and\ \citenamefont
  {Kwiat}}]{Lyons2015}%
  \BibitemOpen
  \bibfield  {author} {\bibinfo {author} {\bibfnamefont {K.}~\bibnamefont
  {Lyons}}, \bibinfo {author} {\bibfnamefont {J.}~\bibnamefont {Dressel}},
  \bibinfo {author} {\bibfnamefont {A.~N.}\ \bibnamefont {Jordan}}, \bibinfo
  {author} {\bibfnamefont {J.~C.}\ \bibnamefont {Howell}}, \ and\ \bibinfo
  {author} {\bibfnamefont {P.~G.}\ \bibnamefont {Kwiat}},\ }\href {\doibase
  10.1103/PhysRevLett.114.170801} {\bibfield  {journal} {\bibinfo  {journal}
  {Phys. Rev. Lett.}\ }\textbf {\bibinfo {volume} {114}},\ \bibinfo {pages}
  {170801} (\bibinfo {year} {2015}{\natexlab{b}})}\BibitemShut {NoStop}%
\bibitem [{\citenamefont {Kofman}\ \emph {et~al.}(2012)\citenamefont {Kofman},
  \citenamefont {Ashhab},\ and\ \citenamefont {Nori}}]{Kofman2012}%
  \BibitemOpen
  \bibfield  {author} {\bibinfo {author} {\bibfnamefont {A.~G.}\ \bibnamefont
  {Kofman}}, \bibinfo {author} {\bibfnamefont {S.}~\bibnamefont {Ashhab}}, \
  and\ \bibinfo {author} {\bibfnamefont {F.}~\bibnamefont {Nori}},\ }\href
  {\doibase http://dx.doi.org/10.1016/j.physrep.2012.07.001} {\bibfield
  {journal} {\bibinfo  {journal} {Physics Reports}\ }\textbf {\bibinfo {volume}
  {520}},\ \bibinfo {pages} {43 } (\bibinfo {year} {2012})},\ \bibinfo {note}
  {nonperturbative theory of weak pre- and post-selected
  measurements}\BibitemShut {NoStop}%
\bibitem [{\citenamefont {Starling}\ \emph {et~al.}(2012)\citenamefont
  {Starling}, \citenamefont {Bloch}, \citenamefont {Vudyasetu}, \citenamefont
  {Choi}, \citenamefont {Little},\ and\ \citenamefont {Howell}}]{Starling2012}%
  \BibitemOpen
  \bibfield  {author} {\bibinfo {author} {\bibfnamefont {D.~J.}\ \bibnamefont
  {Starling}}, \bibinfo {author} {\bibfnamefont {S.~M.}\ \bibnamefont {Bloch}},
  \bibinfo {author} {\bibfnamefont {P.~K.}\ \bibnamefont {Vudyasetu}}, \bibinfo
  {author} {\bibfnamefont {J.~S.}\ \bibnamefont {Choi}}, \bibinfo {author}
  {\bibfnamefont {B.}~\bibnamefont {Little}}, \ and\ \bibinfo {author}
  {\bibfnamefont {J.~C.}\ \bibnamefont {Howell}},\ }\href {\doibase
  10.1103/PhysRevA.86.023826} {\bibfield  {journal} {\bibinfo  {journal} {Phys.
  Rev. A}\ }\textbf {\bibinfo {volume} {86}},\ \bibinfo {pages} {023826}
  (\bibinfo {year} {2012})}\BibitemShut {NoStop}%
\bibitem [{\citenamefont {Boyd}\ and\ \citenamefont
  {Gauthier}(2001)}]{Boyd2001}%
  \BibitemOpen
  \bibfield  {author} {\bibinfo {author} {\bibfnamefont {R.~W.}\ \bibnamefont
  {Boyd}}\ and\ \bibinfo {author} {\bibfnamefont {D.~J.}\ \bibnamefont
  {Gauthier}},\ }\href@noop {} {\emph {\bibinfo {title} {Slow and fast
  light}}},\ \bibinfo {type} {Tech. Rep.}\ (\bibinfo  {institution} {University
  of Rochester Institute of Optics},\ \bibinfo {year} {2001})\BibitemShut
  {NoStop}%
\end{thebibliography}%
\end{document}